\documentclass[aps,prapplied,superscriptaddress,twocolumn]{revtex4-2}
\usepackage{amsmath}
\usepackage{amsfonts}
\usepackage{amssymb}
\usepackage{graphicx}
\usepackage{hyperref}
\usepackage{xcolor}
\usepackage{bm}

\newcommand{\ket}[1]{\ensuremath{\left|{#1}\right\rangle}}
\newcommand{\bra}[1]{\ensuremath{\left\langle{#1}\right |}}
\newcommand{\sinc}{\ensuremath{\mathrm{sinc}}}
\newcommand{\mb}{\mathbf}
\newcommand{\bs}{\boldsymbol}
\newcommand{\Trs}{\ensuremath{\mathrm{Tr}_s}}
\newcommand{\rhospdc}{\ensuremath{\rho_{\mathrm{SPDC}}}}
\newcommand{\rhostim}{\ensuremath{\rho_i^{\mathrm{stim}}}}

\definecolor{darkgreen}{rgb}{0.0, 0.4, 0.0}

\begin{document}

\title{Coincidence free certification and quantification of spatial entanglement\\ with stimulated parametric down conversion}

\author{M. G. Damaceno}
\affiliation{Departamento de F\'{i}sica, Universidade Federal de Santa Catarina, CEP 88040 900, Florian\'{o}polis, SC, Brazil}

\author{G. H. dos Santos}
\email{fisica.gu@gmail.com}
\affiliation{Departamento de F\'{\i}sica, Universidad de Concepci\'on, 160 C Concepci\'on, Chile}
\affiliation{Millennium Institute for Research in Optics, Universidad de Concepci\'on, 160-C Concepci\'on, Chile}

\author{N. Rubiano da Silva}
\email{nara.rubiano@ufsc.br}
\affiliation{Departamento de F\'{i}sica, Universidade Federal de Santa Catarina, CEP 88040 900, Florian\'{o}polis, SC, Brazil}

\author{S. P. Walborn}
\email{swalborn@udec.cl}
\affiliation{Departamento de F\'{\i}sica, Universidad de Concepci\'on, 160-C Concepci\'on, Chile}
\affiliation{Millennium Institute for Research in Optics, Universidad de Concepci\'on, 160-C Concepci\'on, Chile}

\author{P. H. Souto Ribeiro}
\email{p.h.s.ribeiro@ufsc.br}
\affiliation{Departamento de F\'{i}sica, Universidade Federal de Santa Catarina, CEP 88040 900, Florian\'{o}polis, SC, Brazil}

\begin{abstract}
Using stimulated emission, a photon pair source can be characterized by seeding the signal mode with a bright classical beam and measuring the stimulated idler field, thus  replacing two-photon coincidence counting with classical intensity detection. We apply this approach to the continuous transverse spatial degrees of freedom of a down conversion source and show that it is possible to certify spatial entanglement with only intensity measurements. We demonstrate this capability through variance-based entanglement and steering witnesses, as well as the Fedorov ratio. This method is useful for studying entanglement properties of photon pair sources in conditions where alignment and photon counting measurements are difficult and time-consuming.
\end{abstract}

\maketitle

\section{Introduction}

Photon pairs produced by spontaneous parametric down conversion (SPDC) are entangled in their transverse position and momentum, a continuous variable (CV) resource that has underpinned  quantum imaging \cite{RYAN2025100044}, and high dimensional encodings~\cite{malik2026highdimensionalquantumphotonicsroadmap}. The correlations are inherited from the pump beam and nonlinear medium: in the standard angular spectrum picture, the pump field fixes the distribution of the transverse momentum sum while the phase matching function fixes that of the difference, so the twin photons are correlated in position and anticorrelated in momentum~\cite{monken98,walborn2010,Schneeloch2016BiphotonBirthZone}.

Verifying CV entanglement rests on a family of criteria built from the moments of the joint state. At second order, the separability problem is settled for Gaussian states by the Simon criteria \cite{Simon2000}, which is the CV form of the Peres Horodecki criterion~\cite{Peres1996,Horodecki1996}. The Duan variance sum criterion ~\cite{duan2000} and the Mancini Giovannetti Vitali Tombesi (MGVT) product criterion ~\cite{mgvt} express the same content as bounds on the variances of the EPR operators $\bs\rho_1 - \bs\rho_2$ and $\mb q_1+\mb q_2$. Beyond second order, the Shchukin Vogel moment hierarchy~\cite{Shchukin2005} and entropic criteria~\cite{walborn09,Toscano2015} can detect non Gaussian entanglement that variance tests can miss~\cite{gomes09b}. A distinct and stronger statement about quantum correlations is the EPR paradox itself: the Reid criterion \cite{reid1989} bounds the product of \emph{conditional} (inference) variances, and its violation certifies EPR steering ~\cite{Wiseman2007}, which implies non-separability, but not all entangled states present EPR steering. 

All of these tests are normally evaluated from the joint probability distributions, reconstructed by coincidence detection between the signal and idler arms, which is photon starved and slow.
The
transverse degree of freedom is genuinely continuous and
high-dimensional, so characterizing the correlations means sampling a
large two-particle position or momentum space, while the coincidence
rate stays low because every detected pair is a single spontaneous
event. Several strategies have been developed to make this tractable.
The traditional approach raster scans a single-photon detector over
the transverse plane of each arm, or a projective filter (such as a
spatial light modulator) over
a large set of optical modes, building the correlation map point by
point at the cost of long acquisition
times~\cite{howell2004,tasca2008,tasca2009,straupe2011,Giovannini2013,Paul2014,Bavaresco2018}.
Direct measurement of the transverse moments avoids part of
this scan but requires ancillas and a more complex setup \cite{HorMeyll2014}.
Single-photon-sensitive cameras, first electron-multiplying CCDs and more
recently single-photon avalanche diode (SPAD) arrays, instead infer the
joint distribution in parallel across many pixels and have certified
high-dimensional spatial entanglement, though frame rate and detector
noise limit them~\cite{Moreau2012,edgar2015,Ndagano2020,Prasad2024}. Compressed sensing and
adaptive sampling can be used to reduce the number of measurements needed to certify
entanglement in very large Hilbert spaces~\cite{Howland2013,Tonolini2014,Tasca2018,Schneeloch2019}. However, all of
these methods still require coincidence detection of the two-photon field, which is more challenging than simply first-order intensity measurement.

Stimulated parametric down conversion (StimPDC) offers a classical alternative. A bright coherent beam seeds the signal mode of the down conversion crystal and stimulates emission into the idler~\cite{Ou1990,Wang1991pra,wang1991josa}. The resulting bright idler field contains information of the spontaneous biphoton and is measured with ordinary intensity detectors. This type of parametric amplification is the idea behind stimulated emission tomography (SET), which can be used to characterize spectral and polarization correlations in SPDC ~\cite{Liscidini2013,Rozema:15,Fang:16,Keller2022} and, more recently, Gaussian spatial modes and orbital angular momentum~\cite{Xu2024}. Its value is practical: it replaces coincidence counting with classical intensity measurement and inherits the brightness of stimulated over spontaneous emission.

In this work we use StimPDC to characterize the transverse position and momentum correlations through entanglement and steering criteria from idler images alone. The central observation is that, with an appropriately prepared seed, a single stimulated idler image is a conditional distribution. A tightly focused seed beam at the crystal acts as a near point like projection of the signal field in the position basis, so idler field when imaged onto a detector is the position distribution conditioned on the signal: $P(\bs\rho_2|\bs\rho_1)$.  A collimated seed beam acts as a momentum projection on the signal, so the far field idler intensity distribution is proportional to $P(\mb q_2|\mb q_1)$. From such pair of probability distributions we evaluate two CV criteria~\cite{mgvt,reid1989}, both written entirely in idler observables with the seed entering only as the label of which conditional is measured. The first is an inseparability witness, the Mancini, Giovannetti, Vitali, Tombesi (MGVT) product of the sum and difference variances, which is equivalent to the covariance matrix partial transpose and certifies entanglement. The second is a steering witness, the Reid product of conditional variances, which certifies the EPR paradox. We report both products well below their respective bounds, certifying spatial entanglement and EPR steering from the idler arm alone. In addition, we evaluate the Fedorov ratio~\cite{Fedorov2007}, which serves as a quantifier of entanglement for the SPDC biphoton state \cite{Mikhailova2008,Pires2009}.   We show that correlation decreases as the pump waist inside the nonlinear medium is reduced, tracking the source entanglement and confirming that the conditional widths probe the pump shaped spatial correlations. In this case, the nonlinear medium is not changed and the entanglement properties vary only through the pump properties \cite{Law2004}. We treat the realistic case in which the seed is not a perfect position or momentum eigenstate but a focused or a finite width collimated beam, and show that the finite seed width only broadens the measured variances, so both tests are conservative and cannot produce a false positive.

\section{Spatial correlations in SPDC and StimPDC}
\label{sec:iso}

In this section, we recall the isomorphism between the SPDC biphoton wavefunction and the StimPDC idler amplitude, following the standard treatment~\cite{monken98,walborn2010,Schneeloch2016BiphotonBirthZone}. 
Under paraxial and quasi monochromatic approximations, the two photon state generated by SPDC at the crystal plane ($z=0$) is 
\begin{equation}
\ket{\psi}
= \iint d\mb q_1\, d\mb q_2\;
\Phi(\mb q_1,\mb q_2)\,
\hat a_1^\dagger(\mb q_1)\,\hat a_2^\dagger(\mb q_2)\ket{0},
\label{eq:SPDC_state}
\end{equation}
with two-photon angular spectrum

\begin{equation}
\Phi(\mb q_1,\mb q_2) = v(\mb q_1+\mb q_2)\,\gamma(\mb q_1 - \mb q_2),
\label{eq:biphoton_amplitude}
\end{equation}
where $v$ is the pump angular spectrum in the interaction region inside the crystal and $\gamma$ is the phase matching function

\begin{equation}
\gamma(\mb q)=\sinc \!\left(\tfrac{L\Delta k_z}{2}\right)e^{iL\Delta k_z/2},
\label{eq:phase_matching_1}
\end{equation}
with

\begin{equation}
\quad \Delta k_z\simeq\frac{|\mb q_1 - \mb q_2|^2}{2k_p}.
\label{eq:phase_matching_2}
\end{equation}

The subscripts $1,2$ denote the down-converted signal and idler photons, and we use $s,i$ interchangeably with $1,2$. 

In SPDC, projecting the signal onto a mode $v_F(\mb q_1)$ before detection (a spatial  filter in the signal arm) gives the conditional two-photon wavefunction at the crystal~\cite{walborn2010}
\begin{equation}
\phi_F(\mb q_2)=\int d\mb q_1\; v_F(\mb q_1)\,\Phi(\mb q_1,\mb q_2),
\label{eq:phi_cond}
\end{equation}
and the coincidence rate at the idler is $\propto|\phi_F(\mb q_2)|^2$.

In StimPDC, a coherent field with angular spectrum $v_s(\mb q)$ seeds the
signal mode and stimulates a bright idler whose amplitude at $z=0$
is~\cite{Ribeiro1999,Oliveira2020}
\begin{equation}
\phi_i(\mb q_2)=\int d\mb q_1\; v_s^*(\mb q_1)\,\Phi(\mb q_1,\mb q_2),
\label{eq:stim_amp}
\end{equation}
measured directly as the intensity $\mathcal I(\mb q_2)\propto
|\phi_i(\mb q_2)|^2$. Equations~(\ref{eq:phi_cond}) and
(\ref{eq:stim_amp}) coincide under $v_F(\mb q_1)\leftrightarrow
v_s^*(\mb q_1)$: the conditional SPDC idler and the StimPDC idler
amplitude are the same object when the seed is the complex conjugate of
the signal filter. 
 This is
the established basis of SET \cite{Liscidini2013,Rozema:15,Fang:16,Keller2022} in the transverse spatial domain. We note that tomographic reconstruction of continuous variable states is typically much more complicated than the discrete variable case, and is not necessary to determine entanglement.  Instead, here we use the relation of the idler field in StimPDC with the two-photon SPDC amplitude to perform more efficient entanglement tests.

\section{The stimulated idler is the conditional state of the pair}
\label{sec:identity}

The correspondence above is a statement about amplitudes. We now lift it
to density operators, where it becomes the conditional state of the pair.
Write the pure biphoton operator $\rhospdc=\ket\psi\bra\psi$ with matrix
elements $\rho_{spdc}(\mb q_1,\mb q_2;\mb q_1',\mb q_2')
=\Phi(\mb q_1,\mb q_2)\,\Phi^*(\mb q_1',\mb q_2')$, and the seed operator
$\rho_{v_s}=\ket{v_s}\bra{v_s}$ with $\rho_{v_s}(\mb q,\mb q')
=v_s(\mb q)\,v_s^*(\mb q')$. The stimulated idler is the outer product of
the amplitude~(\ref{eq:stim_amp}) with itself,
$\rhostim(\mb q_2,\mb q_2')=\phi_i(\mb q_2)\,\phi_i^*(\mb q_2')$, which
expands to
\begin{equation}
\rhostim(\mb q_2,\mb q_2')
=\iint d\mb q_1\,d\mb q_1'\;
\Phi(\mb q_1,\mb q_2)\Phi^*(\mb q_1',\mb q_2')\,
v_s^*(\mb q_1)\,v_s(\mb q_1').
\label{eq:expand}
\end{equation}
Using $v_s^*(\mb q_1)\,v_s(\mb q_1')=\rho_{v_s}(\mb q_1',\mb q_1)$, Eq.~(\ref{eq:expand}) becomes
\begin{equation}
\rhostim(\mb q_2,\mb q_2')
=\iint d\mb q_1\,d\mb q_1'\;
\rhospdc(\mb q_1,\mb q_2;\mb q_1',\mb q_2')\,
\rho_{v_s}(\mb q_1',\mb q_1),
\label{eq:expand2}
\end{equation}
which is the partial trace over the signal of $\rhospdc$ with the seed
operator inserted on the signal factor,
\begin{equation}
\rhostim
=\Trs\!\big[\rhospdc\,(\rho_{v_s}\otimes I_i)\big]
=\bra{v_s}\rhospdc\ket{v_s}_s .
\label{eq:central}
\end{equation}
This is exactly the spontaneous conditional idler state that a
coincidence measurement projecting the signal onto $\ket{v_s}$ would
herald, now read out as a bright classical intensity.

\section{Position and momentum seeds}
\label{sec:signflip}

\begin{figure}
    \includegraphics[width=1\linewidth]{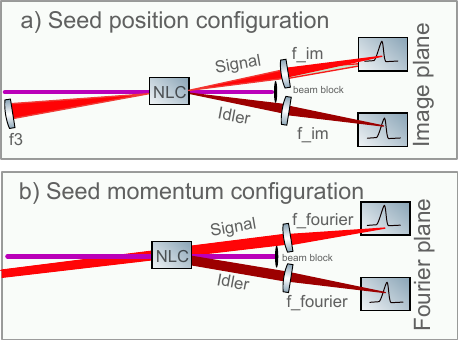}
\caption{Conceptual diagram of the two seed beam configurations used to obtain the conditional distributions of the idler beam. Only first-order intensity measurements are performed. Measurements of the signal field are used only to calibrate the seed profile at the crystal. Details in the main text.}
\label{fig:seed_configs}
\end{figure}

Two limiting seed fields provide the required conditional distributions used in the
entanglement test. The experimental scheme for the case where the nonlinear medium is a bulk crystal is illustrated in Fig.~\ref{fig:seed_configs}. 
The seed beam (bright red) overlaps with the pump beam (magenta) in the nonlinear crystal (NLC). The amplified seed emerges as the signal beam (bright red), used to monitor the seed profile, while the stimulated emission emerges as the idler beam (dark red). In Fig.~\ref{fig:seed_configs}a) the position-space seed configuration is shown. An additional lens ($f_3$) focuses the seed to a small spot (approximately point like) within the crystal. This position seed $u(\mb r)\propto\delta(\mb r - \mb r_0)$ gives 
\begin{equation}
\rhostim = \bra{\mb r_0}\rhospdc\ket{\mb r_0}_s,
\end{equation}
the conditional idler given the signal at $\mb r_0$. The signal/idler outputs are projected onto the image plane ($f_{\mathrm{im}}$), resulting in $P(\bs\rho_2|\bs\rho_1=\mathbf r_0)$. In Fig.~\ref{fig:seed_configs}b) momentum space seed configuration is shown. The seed is collimated (no focusing lens). This momentum seed $u(\mb r)\propto e^{i\mb q_0\cdot\mb r}$ gives
\begin{equation}
\rhostim = \bra{\mb q_0}\rhospdc\ket{\mb q_0}_s,
\end{equation}
the conditional idler given the signal momentum $\mb q_0$. Projecting the outputs onto the Fourier plane
($f_{\mathrm{fourier}}$) gives $P(\mathbf q_2|\mathbf q_1=\mathbf q_0)$. 
For a
momentum anticorrelated source ($\Phi\propto\delta(\mb q_1+\mb q_2)$) in
the thin crystal/EPR limit) the idler then peaks at $  \mb q_0$, the
anticorrelated partner, which is the same result a coincidence projection
of the signal onto $\mb q_0$ would give. StimPDC reproduces the
spontaneous conditional idler in both bases, now as a bright classical
image. These near and far field conditional distributions are the inputs
to the entanglement witness of the next section.  Note that here we considered that the seed is prepared in perfect position or momentum eigenstates, finite width effects will be considered in the next section.

\section{Idler-only entanglement tests}
\label{sec:mancini}
Because each idler image is a conditional distribution, the natural entanglement witness is built from conditional variances, the same
inference variances that enter the EPR and steering
criteria~\cite{reid1989,Wiseman2007}. The Reid criteria is given by
the product of conditional variances
\begin{equation}
\mathcal{W}
= \Delta^2(\bs\rho_2|\bs\rho_1)\,\Delta^2(\mathbf q_2|\mathbf q_1) \geq \frac{1}{4}.
\label{eq:witness}
\end{equation}
A non-steerable state obeys
the above inequality, while a measured
$\mathcal{W}<1/4$ certifies quantum steering and also entanglement~\cite{Wiseman2007}. In StimPDC both factors are idler-only observables conditioned on preparation of the seed beam, so
Eq.~(\ref{eq:witness}) is a test on the signal  idler pair
that never detects the signal photon and never reconstructs the joint
distribution.

The two conditional widths also probe the source structure. With
$\Phi(\mathbf q_1,\mathbf q_2)=v(\mathbf q_1+\mathbf q_2)\gamma(\mathbf q_1 - \mathbf q_2)$, the conditional momentum width is set by the pump
angular spectrum $v$, narrow for a collimated pump, while the conditional
position width is set by the phase matching function $\gamma$ and the pump
profile at the crystal. The product therefore measures how the pump and
crystal shape the spatial correlations, and tightening the pump (Sec. ~\ref{sec:results}) tunes the entanglement directly.

The MGVT criterion certifies entanglement when~\cite{mgvt}
\begin{equation}
\Delta^2(\bs\rho_1 - \bs\rho_2)\cdot\Delta^2(\mb q_1+\mb q_2) < 1.
\label{eq:mgvt}
\end{equation}
Previous experiments have used this criterion, as well as similar ones using higher-order moments \cite{Shchukin2005} or entropy functions \cite{walborn09}, on the photon pairs produced from SPDC, which requires joint detection of the signal and idler fields in their transverse planes \cite{tasca08,gomes09b,Schneeloch2019,Ndagano2020}.  
 Equation~(\ref{eq:central}) removes the signal arm entirely, evaluating
bipartite entanglement criteria
accessible by the capture of a sequence of images of the idler field for different preparations of the seed. 
Seeding with a position eigenstate at $\mb r_0$ gives the exact
conditional idler distribution $P(\bs\rho_2\,|\,\bs\rho_1=\mb r_0)$.
Scanning $\mb r_0$ over the crystal plane and stacking the idler images,
weighted by the signal marginal $P(\bs\rho_1)$, synthesizes the full
joint distribution $P(\bs\rho_1,\bs\rho_2)$; the variance
$\Delta^2(\bs\rho_1 - \bs\rho_2)$ is then a single moment of this measured
distribution. The momentum term follows identically from a far field
scan, with the seed prepared as a plane wave carrying a linear phase that
defines $\mb q_0$. Seeding momentum $\mb q_0$ heralds the idler at its
anticorrelated partner, so the synthesized joint momentum distribution is
peaked along $\mb q_1+\mb q_2\approx0$ and $\Delta^2(\mb q_1+\mb q_2)$ is
small for an entangled source. The MGVT
product~(\ref{eq:mgvt}), or any entanglement criteria dependent on the joint position and momentum distributions \cite{Shchukin2005,walborn09,Toscano2015}, can thus be reconstructed from idler intensity images
alone.  Importantly, the intensity of the stimulated idler field is such that it can typically be measured using a simple and inexpensive CCD camera.  

\paragraph*{Finite seed width.}
A real seed is neither a point nor a plane wave, so each measurement uses
the seed in the basis where it is sharp: a focused beam (small position
width $\sigma_\rho$) for the position conditional, a collimated beam (small
momentum width $\sigma_q$) for the momentum measurement. A
finite width averages the conditional idler distributions over a range of seed values given by these widths and
only broadens the measured variance,
\begin{equation}
\Delta^2_{\mathrm{meas}}
= \Delta^2_{\mathrm{true}} + \sigma_{\mathrm{seed}}^2 ,
\label{eq:bias}
\end{equation}
with $\sigma_{\mathrm{seed}}$ the seed width in the measured variable
($\sigma_\rho$ for position, $\sigma_q$ for momentum). The bias is positive and can be calibrated: one can measure the seed profile in each configuration and subtract it from $\Delta^2_{\mathrm{meas}}$.
Because it always \emph{increases} both factors of the entanglement and steering witnesses we will be evaluating,
an uncorrected violation is conservative, so that the bias can only underreport
entanglement and never produce a false positive \cite{Tasca2013}. The complementary spreads
of the two seeds do not conflict, since each seed is used only for the
conjugate variable in which it is narrow.

\section{Experimental setup}
\label{sec:setup}

\begin{figure}[h]
\centering
\includegraphics[width=\linewidth]{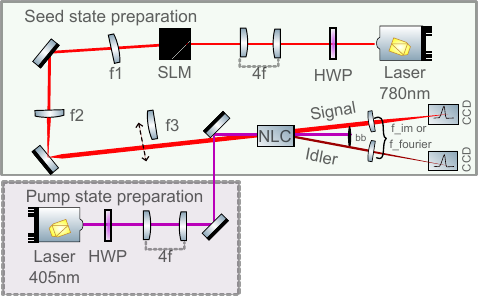}
\caption{Experimental setup. See main text for details.}
\label{fig:setup}
\end{figure}

Figure~\ref{fig:setup} presents a sketch of the experimental setup.  The pump beam is a $405$~nm laser,
vertically polarized and collimated by a 4$f$ system. The seed beam is prepared independently using a horizontally polarized $780$~nm diode laser that is shaped into a good quality Gaussian mode using a spatial light modulator (SLM). 
Lenses $f_1=100$~mm and $f_2=200$~mm form a telescope with magnification $M=2$, preparing a collimated seed beam inside the crystal. Pump and seed beams
are overlapped inside a 2 mm thick type I BBO crystal at a crossing
angle of approximately $4.5^\circ$, satisfying the phase matching
condition. Energy conservation yields the stimulated idler
wavelength giving $\lambda_i=840$~nm  for
$\lambda_p=405$~nm and $\lambda_s=780$~nm.

After the crystal, light from the pump is removed with a beam block before the imaging optics. The image plane and Fourier plane measurements of the seed and idler fields are selected by the final lens ($f_{\mathrm{im}}=150$~mm or $f_{\mathrm{Fourier}}=300$~mm), implementing the position and momentum measurements described in Secs.~\ref{sec:signflip} and \ref{sec:mancini} with magnification equal to 1.  Two CMOS cameras simultaneously record the stimulated idler and the seed beam, the latter serving only as a calibration reference. No coincidence counting is performed.
The cameras were equipped with band-pass filters with a band of $\pm10$nm. For the momentum configuration, the convention used to convert the position in the Fourier plane to momentum was $q= 2 \pi y/\lambda f_{\text{fourier}} $.

Two seed configurations are employed. For the {\it position} configuration, the seed is focused inside the crystal using a lens ($f_3=400$~mm)  placed with
the crystal located in its focal plane, focusing the seed into an
approximately point-like spot.   The {\it momentum} configuration is realized by collimating the seed inside the crystal (lens $f_3$ is removed). 
 Both $f_1$ and $f_3$ are mounted on
translation stages, allowing controlled scans of the seed momentum
and transverse position, respectively.

\section{Results}
\label{sec:results}

We perform intensity measurements for both preparations of the seed beam.
 From the camera images, we obtain marginal intensity distributions along the vertical laboratory direction $y$ (with respect to the surface of the optical table) by averaging each image along the perpendicular $x$ axis. Six different seed positions and momenta are scanned across the $y$ direction, and 10 samples are captured for each of them. From these 1D profiles (intensity $I$ vs. position $\rho_y$ or momentum $q_y$), we calculate the quantities used for all entanglement and steering tests.   

\begin{figure}
    \centering
    \includegraphics[width=
    \linewidth]{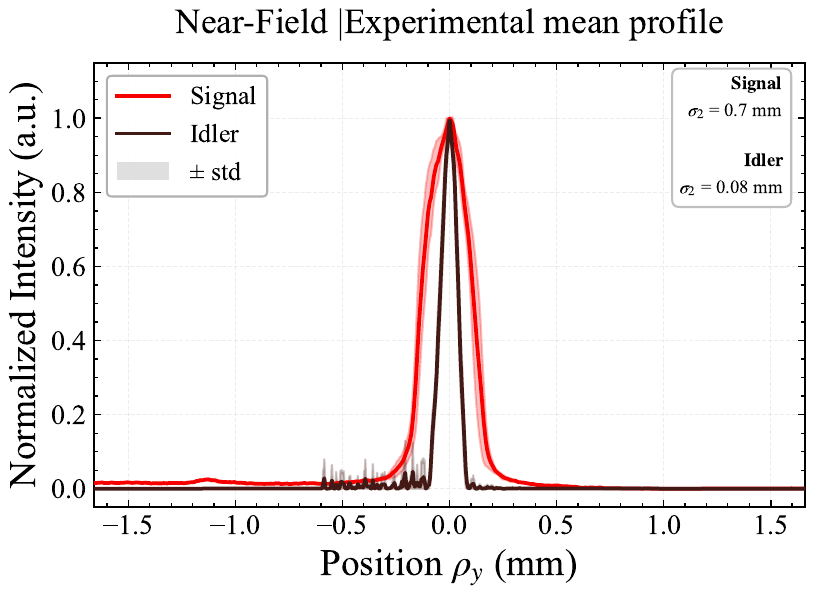}
    \caption{ Near-field intensity profile along the vertical position for the position configuration (image plane): signal and stimulated idler mean profiles over $N=10$ repetitions; shaded bands indicate the standard deviation across repetitions. Second moment widths: $\sigma_2=0.7$~mm (signal), $\sigma_2=0.08$~mm (idler).}
    \label{fig:near_field}
\end{figure}

\begin{figure}
    \centering
    \includegraphics[width=\linewidth]{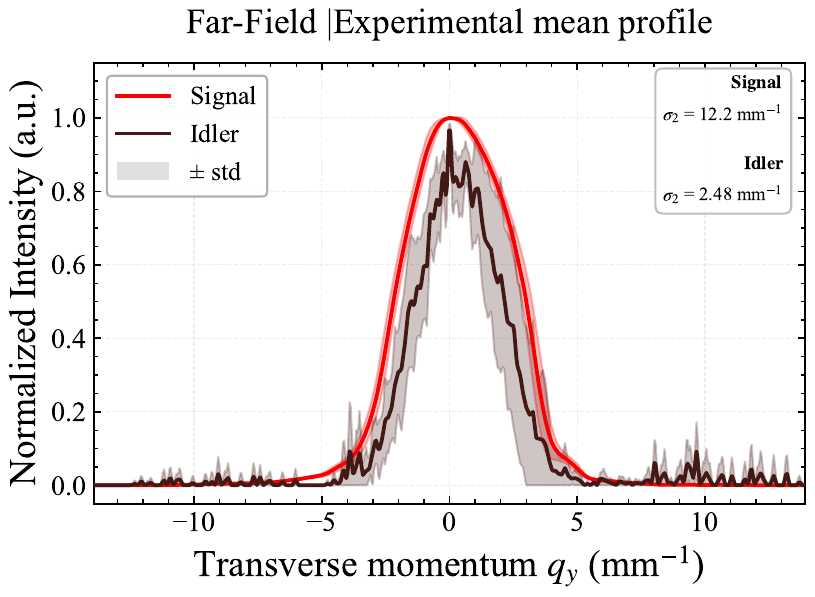}
    \caption{Far-field intensity profile vs. transverse, vertical momentum $q$, for the momentum configuration (Fourier plane): seed (signal) and stimulated idler mean profiles over $N=10$ repetitions; shaded bands indicate the standard deviation across repetitions. Second moment widths: $\sigma_2=12.2$~mm$^{-1}$ (signal), $\sigma_2=2.48$~mm$^{-1}$ (idler).}
    \label{fig:far_field}
\end{figure}

Figs. \ref{fig:near_field} and \ref{fig:far_field} show examples of position and momentum intensity distributions we measured, along $\rho_y$ and $q_y$, respectively. For each distribution, the variances conditioned to the seed configuration are evaluated from the model-independent second moment
\begin{equation}
\sigma_2 = \sqrt{\frac{\sum_a (a- \bar{a})^2\,I(a)}{\sum_a I(a)}},
\quad
\bar{a} = \frac{\sum_a a\,I(a)}{\sum_a I(a)},
\label{eq:sigma2_def}
\end{equation}
where $a$ is equal to $\rho_y$ ($q_y$) for the position (momentum) configuration.
No functional form for the measured intensity distribution is assumed.


\begin{figure}
    \centering
    \includegraphics[width=\linewidth]{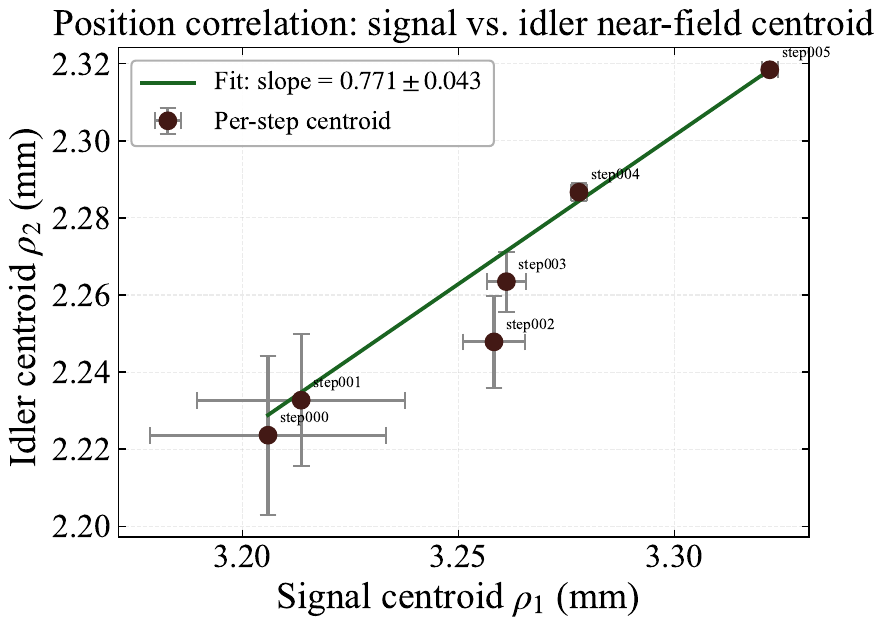}
        \caption{Position correlation: idler near field centroid
        $\rho_2$ vs.\ signal near field centroid $\rho_1$ across
        the six seed positions. Linear fit slope
        $0.77\pm0.04$.}
        \label{fig:corr_pos}
\end{figure}
\begin{figure}
    \centering
    \includegraphics[width=\linewidth]{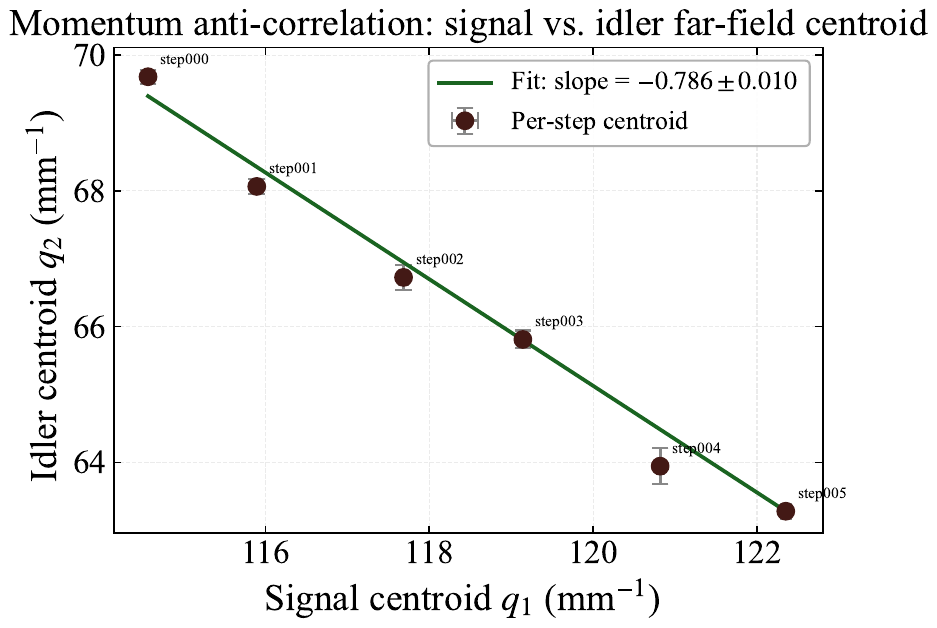}
    \caption{Momentum anticorrelation: idler far field centroid
        $q_2$ vs.\ signal far field centroid $q_1$. Linear fit
        slope $ 0.79\pm0.01$, confirming anticorrelation.}
        \label{fig:corr_mom}
\end{figure}

Figure \ref{fig:corr_pos} (Fig. \ref{fig:corr_mom}) shows the relationship between the seed and idler position (momentum) inside the crystal for six different values. The coordinates of each point are obtained from the centroid of the position (momentum) distribution obtained by imaging (Fourier transforming) seed and idler beams. Position (momentum) correlation (anticorrelation) can be clearly observed.  

Linear fits give slopes of $0.77\pm0.04$ for the position and $- \,0.79\pm0.01$ for the momentum measurements, confirming the expected position correlation and momentum anticorrelation. The deviation from the
ideal slopes $\pm 1$ is related to the different seed and idler wavelengths and other technical issues affecting the calibration of the length scales. In any case, what matters for the correlations is the dispersion of the points around the linear fit. This dispersion is quantified by the fitting error bars. Imperfections that broaden the measured distributions can only increase the evaluated witnesses and cannot produce false certifications of entanglement and steering.

Let us analyze the steering witness by using the measured conditional widths for each position (momentum) measurement displayed in Fig. \ref{fig:corr_pos} (Fig. \ref{fig:corr_mom}). The results are listed in Table~\ref{tab:reid_steps}, together with the corresponding Reid steering witness given by Eq. \eqref{eq:witness} written in terms of the measured variances $\sigma_{2,\mathrm{IM}}$ position configuration and $\sigma_{2,\mathrm{MO}}$ momentum
\[
\mathcal W
=
\sigma_{2,\mathrm{IM}}^2\, \sigma_{2,\mathrm{MO}}^2 \, .
\]

\begin{table}[ht]
\centering
\caption{Per step conditional widths of the idler beam and Reid steering witness $\mathcal{W}$.}
\label{tab:reid_steps}
\begin{tabular}{cccc}
\hline
Step & $\sigma_{2,\mathrm{IM}}$ (mm) & $\sigma_{2,\mathrm{MO}}$ (mm$^{ 1}$) & $\mathcal{W}$ \\
\hline
0 & $0.08\pm0.01$& $2.53\pm0.17$ & $0.04\pm0.01$ \\
1 & $0.08\pm0.01$ & $2.45\pm0.10$ & $0.04\pm0.01$ \\
2 & $0.08\pm0.01$ & $2.48\pm0.10$ & $0.04\pm0.01$ \\
3 & $0.08\pm0.01$ & $2.52\pm0.13$ & $0.04\pm0.01$ \\
4 & $0.09\pm0.01$ & $2.72\pm0.15$ & $0.05\pm0.01$ \\
5 & $0.09\pm0.01$ & $3.01\pm0.28$ & $0.08\pm0.02$ \\
\hline
Mean & & & $0.05\pm0.02$ \\
$\mathcal{B}$ & & & $0.25$ \\
\hline
\end{tabular}
\end{table}

We can see that all measured values remain well below the bound $\mathcal B=1/4$. Considering the six values, we obtain the average and standard deviation
\[
\mathcal W
=
0.05
\pm
0.02 \,.
\]
The average is approximately five times smaller than the steering limit. Table~\ref{tab:reid_steps} also shows a systematic increase of both variances and $\mathcal W$ from step $0$ to $5$. This is related to the experimental arrangement for scanning position and momentum. However, even the larger variances are narrow enough to ensure the witness is well below the limit. Therefore, this is a consistent and robust verification of EPR steering. 

To verify that the observed steering is consistent with a standard inseparability criterion, we reconstruct the variances entering the Mancini Simon witness using the law of total variance,
\begin{equation}
\Delta^2(\rho_1 - \rho_2)
= \underbrace{\bigl\langle\Delta^2(\rho_2|\rho_1)\bigr\rangle_{P(\rho_1)}}_{\text{Term A}}
+ \underbrace{\mathrm{Var}\bigl(\bar\rho_1^{(k)} \bar\rho_2^{(k)}\bigr)_k}_{\text{Term B}},
\label{eq:totalvar}
\end{equation}
where the first term is obtained directly from the conditional widths of Table~\ref{tab:reid_steps} and the second from the centroid correlations shown in Figs.~\ref{fig:corr_pos} and
\ref{fig:corr_mom}. The centroid variances (Term B) 
are more than one order of magnitude smaller than the conditional variances (Term A) in Eq. \eqref{eq:totalvar}. Therefore Term A dominates the total variance. In order to obtain the error bar for the variance, we perform bootstrap resampling of momenta and positions ($4000$ realizations) yielding
\begin{equation}
\Delta^2(\rho_1 - \rho_2)\cdot\Delta^2(q_1+q_2)
= 0.05\pm0.01.
\label{eq:mancini_result}
\end{equation}
This result, well below the MGVT bound of 1 in Eq. \eqref{eq:mancini_result}, confirms the spatial entanglement through the MGVT inseparability witness reconstructed from the measured conditional and centroid variances. 

\begin{figure}
    \centering
    \includegraphics[width=\linewidth]{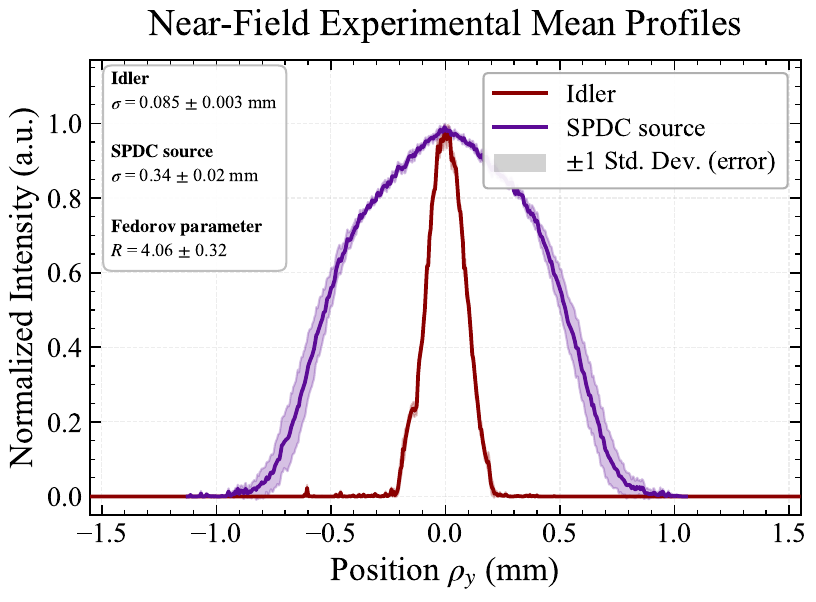}
    \caption{Comparison between the stimulated and spontaneously emitted idler intensity profiles for a collimated pump. Insets give the second momentum widths and the Fedorov parameter
$R=\sigma_{\mathrm{SPDC}}/\sigma_{\mathrm{Stim}}$: $R=4.06\pm0.32$. }
    \label{fig:fedorov_coli}
\end{figure}

\begin{figure}
    \centering
    \includegraphics[width=\linewidth]{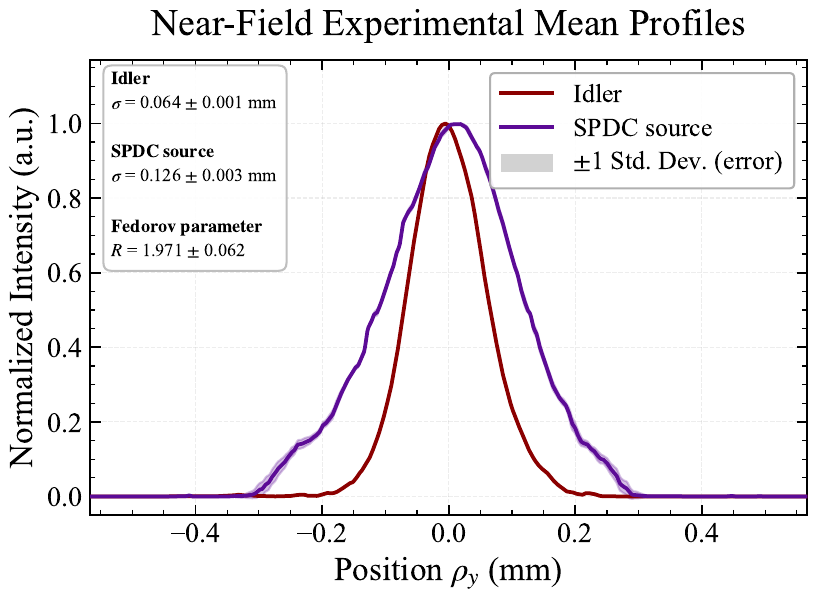}
    \caption{Comparison between the stimulated and spontaneously emitted idler intensity profiles with a focused pump at the crystal plane. Insets give the second momentum widths and the Fedorov parameter $R=1.97\pm0.06$ (focused using a lens of $f=750$~mm).}
    \label{fig:fedorov_foc}
\end{figure}

Finally, we characterize the spatial entanglement using the Fedorov ratio \cite{Fedorov2007}, which is a fair quantifier for pure states,
\begin{equation}
    R
=
\frac{\sigma_{\mathrm{SPDC}}}
{\sigma_{2,\mathrm{IM}}},
\label{eq:fedorov_ratio}
\end{equation}

where the unconditioned SPDC is measured through the spontaneous emitted idler light, without seeding. A separable state gives $R=1$, so $R>1$ indicates entanglement, and in fact $R$ is equivalent to the Schmidt number for gaussian states \cite{Mikhailova2008}. Knowing that the pump beam width inside the crystal affects the entanglement, we test two configurations, collimated and focused pump.  For the collimated pump we obtain $R=4.06\pm0.32$, whereas for the focused pump, the ratio reduces to $R=1.97\pm0.06$.
The measured intensity distributions are shown in Figs.~\ref{fig:fedorov_coli} and ~\ref{fig:fedorov_foc}. This behavior is consistent with the
reduction of the effective Schmidt number as the pump becomes more
tightly focused, providing an independent confirmation of the
pump-controlled spatial correlations.

The results shown here have a practical and a conceptual aspect. Practically, it turns a two photon entanglement test into a single arm classical measurement of two idler images, inheriting the brightness advantage of stimulated over spontaneous emission, valuable where coincidence counting is slow: faint pumps, weak nonlinear interaction, restricted spatial resolution imposed by the tradeoff between small detector collection area and photon counting and coincidence rate, low image capture rate in the case of single-photon cameras, and low quantum efficiency of photon counters for some wavelength ranges. 

Conceptually, the conditional widths demonstrate how the pump and crystal specifications shape the spatial correlations. Moreover, as noted in Sec.~\ref{sec:identity}, the same idler data may be read as a contraction of the partially transposed biphoton operator, linking the measurement to the separability criteria.

\section{Conclusion}

We have used stimulated emission in a three-wave nonlinear interaction to certify entanglement and steering of transverse spatial
position and momentum degrees of freedom of photon pairs spontaneously emitted in the process directly from idler intensity measurements instead of photon coincidence. A focused or a collimated seed turns each idler intensity pattern into a conditional distribution. The conditional variances $\Delta^2(\bs\rho_2|\bs\rho_1)$ and $\Delta^2(\mathbf q_2|\mathbf q_1)$ are read off from two intensity distributions, and their
product is an entanglement witness written entirely in idler observables,
with the seed serving only as a label. 

Our results show a product well below the bound for an EPR-type certification of spatial entanglement and steering, and a
Fedorov ratio quantifies entanglement as for two different pump waists  so that the conditional widths lead to large and small entanglement strength. Practical issues like the finite width of a realistic focused or collimated seed beam only broaden the measured variances, so the test is conservative.  Reading an entanglement witness off from a bright
beam, rather than from coincidences, is a practical advantage where photon
counting is difficult. 

\begin{acknowledgements}
This work has been supported by the Brazilian agencies CNPq
(DOI 501100003593), CAPES (DOI 501100002322), FAPESC
(DOI 501100005667, DOI 2025TR001683), and INCT IQ (465469/2014 0),
INCT IQNano (406636/2022 2), INCT DQ (408783/2024 9); and by the Chilean Fondo Nacional de Desarrollo
Cient\'{\i}fico y Tecnol\'ogico (FONDECYT) Grant No.\ 1240746,
ANID -- Millennium Science Initiative Program -- ICN17$\_$012,
and ANID Anillo Project ATE250003.
\end{acknowledgements}



\bibliographystyle{apsrev4-2}
\bibliography{biblio.bib}

\end{document}